\documentclass[aps,prb,twocolumn,superscriptaddress]{revtex4-1}
\usepackage{epsfig}
\usepackage{times}
\usepackage{color}
\usepackage{graphicx}
\usepackage{float}
\usepackage{amsmath}
\usepackage[caption=false]{subfig}

\begin{document}

\title{Inversion of the chemical environment representations}
\author{Matteo Cobelli}
\affiliation{School of Physics, AMBER and CRANN Institute, Trinity College, Dublin 2, Ireland}
\author{Paddy Cahalane}
\affiliation{School of Physics, AMBER and CRANN Institute, Trinity College, Dublin 2, Ireland}
\author{Stefano Sanvito}
\affiliation{School of Physics, AMBER and CRANN Institute, Trinity College, Dublin 2, Ireland}

\begin{abstract}
Machine-learning generative methods for material design are constructed by representing a 
given chemical structure, either a solid or a molecule, over appropriate atomic features, generally 
called structural descriptors. These must be fully descriptive of the system, must facilitate the training 
process and must be invertible, so that one can extract the atomic configurations corresponding to the 
output of the model. In general, this last requirement is not automatically satisfied by the most efficient 
structural descriptors, namely the representation is not directly invertible. Such drawback severely limits 
our freedom of choice in selecting the most appropriate descriptors for the problem, and thus our flexibility 
to construct generative models. 
In this work, we present a general optimization method capable of inverting any local many-body descriptor 
of the chemical environment, back to a cartesian representation. The algorithm is then implemented 
together with the bispectrum representation of the local structure and demonstrated for a number of
molecules. The scheme presented here, thus, represents a general approach to the inversion of
structural descriptors, enabling the construction of efficient structural generative models.
\end{abstract}

\maketitle


Machine-learning (ML) algorithms are becoming increasingly popular as an alternative to conventional
electronic structure theory in condensed matter physics. Their most successful application is perhaps
in the construction of highly accurate machine-learning atomic potentials (MLAPs). These are often 
trained to reproduce labelled data obtained with {\it ab-initio} methods, such as density functional 
theory (DFT), at a fraction of the computational cost.\cite{Behler-Parrinello,Gap,snap,ANI,gdml} 
Their accuracy can reach that of the electronic-structure theory used to construct the training data 
and it is essentially limited by the volume of data used for the training.\cite{Lunghi2019a} Furthermore, 
the same models can be used to predict properties different from the total energy.\cite{Lunghi2019b}

One of the key aspects in the creation of a ML model is the definition of a suitable set of input 
features that fully describes the data,\cite{Musil2021} namely the atomic and chemical structure of a 
solid or a molecule. Such features, usually called {\it structural descriptors}, should be as descriptive 
as possible and should facilitate the training process of the given ML model. In the case of MLAPs 
the inclusion of system symmetries in the definition of the descriptors has proven to substantially 
improve the performance of the models and to reduce the amount of data needed for the 
training.\cite{Behler-Parrinello,Bartok} Crucially, in the construction of MLAPs the structural descriptors 
do not necessarily need to be readily interpretable. This means that one has to associate to a given
structure a set of descriptors, but never has to answer the inverse question, namely which structure
is associated to a set of descriptors, when this is given. 

The same is not true for generative methods, which are algorithms that produce prototype structures
according to given distributions.\cite{Sanchez-Lengeling2018} In this case, in fact, the ML algorithm 
works with a molecular representation based on structural descriptors, but the output should
be an interpretable structure, for instance the chemical identity and cartesian coordinates of the atoms 
forming a molecule. As such, the molecular representation should also be invertible. Possible solutions to
this problem include representations that distinguish between different structures based on the concept 
of chemical bonds, such as the SMILES encoding for organic molecules\cite{smiles1,smiles2} or general
graphs encoding.\cite{drugan,molgan}. These methods, by construction, are capable to capture the general
structure of a chemical entity, but they cannot distinguish between different deformations of the same molecule.
For instance, all the configurations encountered by a given molecule over a molecular dynamic trajectory 
will share the same encoding.

As a solution, one can construct representations based on fractional coordinates with respect to a unit 
cell, which are then able to distinguish between distortions of the same system. These, however, 
lack of rotational and translational invariance and heavily rely on data augmentation to incorporate the 
fundamental symmetries in the model.\cite{crystal_gan_3} Alternatively, one can constrain the problem 
to a very specific family of structures and discretise the possible atomic positions, so that the inversion
from a given representation to the cartesian coordinates and the model training is more easily 
achieved.\cite{crystal_gan_2,crystal_gan_1} This last strategy, however, lacks of universality.

Clearly, the ideal solution to all these issues would be that of developing a general algorithm to invert 
the structural descriptors used for constructing MLAPs back to a cartesian representation. Our paper 
presents such a method. In particular, we have built a general scheme to invert any representation 
based on many-body local structural descriptors, which then can be used in any generative algorithm. 
Here, we introduce our general method, with its most relevant numerical details, and show examples 
for a specific MLAPs representation, namely the bispectrum.\cite{Bartok}

\begin{figure*}[t]
\centering
\includegraphics[scale=0.5]{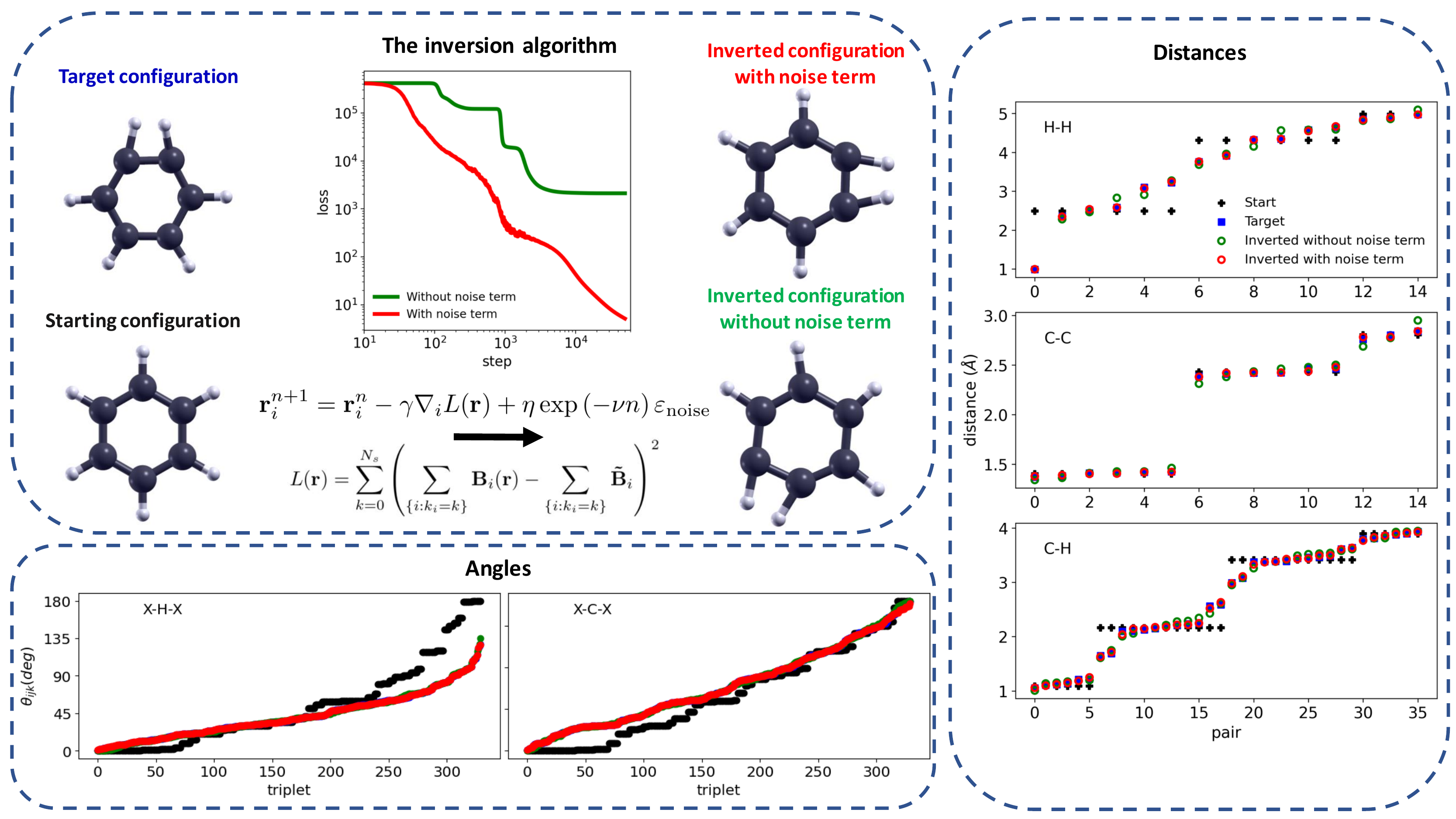}
\caption{(Color on-line) Scheme associated to an example of inversion of the bispectrum components starting 
from a relaxed benzene structure. The target is a deformed benzene molecule chosen for sake of visualization. 
A comparison of the collection of distances and planar angles from all the atomic pairs and triplets in the 
molecule are shown. After $5\cdot10^{4}$ iterations of the gradient descent algorithm, the inverted 
configuration (red) closely resembles the target one (blue). We also show in green the result of the inversion 
process in the absence of the noise term in the update rule of Eq.~(\ref{update_rule}), see text for details.}
\label{fig:inversion_scheme}
\end{figure*}
%
%
Most MLAPs assume that the total energy of a molecule/solid can be expressed as a sum of atomic 
contributions, in turn, depending on the local environment of each atom of the system. For instance, 
the total energy of a molecule made of $N$ atoms can then be written as,\cite{Behler-Parrinello,Bartok,snap}
\begin{equation}\label{total energy}
        E_\mathrm{tot} =\sum_{i=0}^{N}\varepsilon_{i}(\mathbf{B}_{i}),
\end{equation}
where the energy, $\varepsilon_{i}$, associated to the atom at the position $\mathbf{r}_{i}$, is a function 
of the descriptors of the atomic environment, $\mathbf{B}_{i}$. These are then function of the position 
of all the atoms within a cutoff radius, $r_\mathrm{c}$, from $\mathbf{r}_{i}$. The choice of the specific 
descriptors significantly impacts the performance of the MLAP. As such, it is often necessary to construct 
the $\mathbf{B}_{i}$'s so to satisfy the symmetries of the quantity that one wants to predict. In the case of 
the total energy, the structural descriptors are designed to be invariant with respect to rotations, while the 
form of Eq.~(\ref{total energy}) guarantees that $E_\mathrm{tot}$ remains invariant against translation 
and atomic permutations. As a result, the best performing local descriptors are often many-body in 
nature\cite{Musil2021} and their transformation from the Cartesian coordinates is not globally invertible.

Here, we show that the local inversion of this transformation can be achieved through the optimisation of 
an initial atomic configuration, by means of a gradient descent algorithm. The main idea, see 
Fig.~\ref{fig:inversion_scheme}, consists in optimising a molecular structure so that its descriptors 
representation matches the given set of target descriptors (for instance, obtained from a generative model). 
Thus, given a set of target descriptors, $\{\mathbf{\tilde{B}}_{i}\}$, and the Cartesian coordinates of a starting 
configuration, $\{\mathbf{r}_{i}\}$, our algorithm updates the atoms positions until the associated structural 
descriptors of the molecule, $\{\mathbf{B}_{i}(\mathbf{r})\}$, coincide with $\{\mathbf{\tilde{B}}_{i}\}$ within 
a numeric tolerance. The distance between the target descriptors and the optimised ones can be quantitatively 
measured via a loss function,
\begin{equation}\label{inversion_loss}
    L(\mathbf{r}) =\frac{1}{N_\mathrm{s}}\sum_{k=0}^{N_s}\left( \sum_{\{i:k_{i}=k\}} \mathbf{B}_{i}(\mathbf{r})- \sum_{\{i:k_{i}=k\}} \mathbf{\tilde{B}}_{i}\right)^{2}\:,
\end{equation}
where $N_\mathrm{s}$ is the number of distinct chemical species present in the system, so that the external
sum runs over the possible species and the internal one over the atoms belonging to a given specie. The form 
of $L(\mathbf{r})$ has been chosen to be invariant under permutation of atoms of the same species. By using 
a gradient descend algorithm\cite{Bishop} it is then possible to update the coordinates of the initial configuration 
so to minimise $L(\mathbf{r})$. At the $n$-th iteration the $(n+1)$-th update of the Cartesian coordinates 
$\{\mathbf{r}_i\}$ is given by,
\begin{equation}\label{update_rule}
    \mathbf{r}_{i}^{n+1} = \mathbf{r}_{i}^{n} - \gamma \nabla_{i}L(\mathbf{r}) +\eta\exp{\left(-\nu n\right)\varepsilon_\mathrm{noise}}\:, 
\end{equation}
where $\gamma$ is the learning rate and $\varepsilon_\mathrm{noise}\in[0,1]$ is a random number generated 
at each gradient-descend iteration. The second term in Eq.~(\ref{update_rule}) is introduced to break the symmetry
at configurations where the gradients of the descriptors of the local environment tend to vanish and, in general,
it is found to stabilise the inversion process. The coefficient $\eta$ determines the coupling strength and $\nu$ 
controls its exponential decay with the iteration number.

Since the relation between the Cartesian coordinates and the atomic descriptors is not invertible, the loss function
of Eq.~(\ref{inversion_loss}) has multiple global minima. For instance, for a set of rotationally invariant descriptors, 
every rotation of the target configuration will correspond to a global minimum with $L=0$. However, the region of 
the possible coordinates explored during the optimisation process is limited by the choice of the initial configuration. 
This guarantees the possibility of local inversion of the relation between the Cartesian coordinates and the descriptors. 
Note that the same argument applies when the descriptors are incomplete\cite{Incompleteness}, and therefore a further 
increase in the number of global minima of $L(\mathbf{r})$ is expected. As such, in general, the configuration reached 
at convergence will depend on the initial configuration, which then needs to be cleverly chosen.


\begin{figure}
\centering
\includegraphics[scale=0.40]{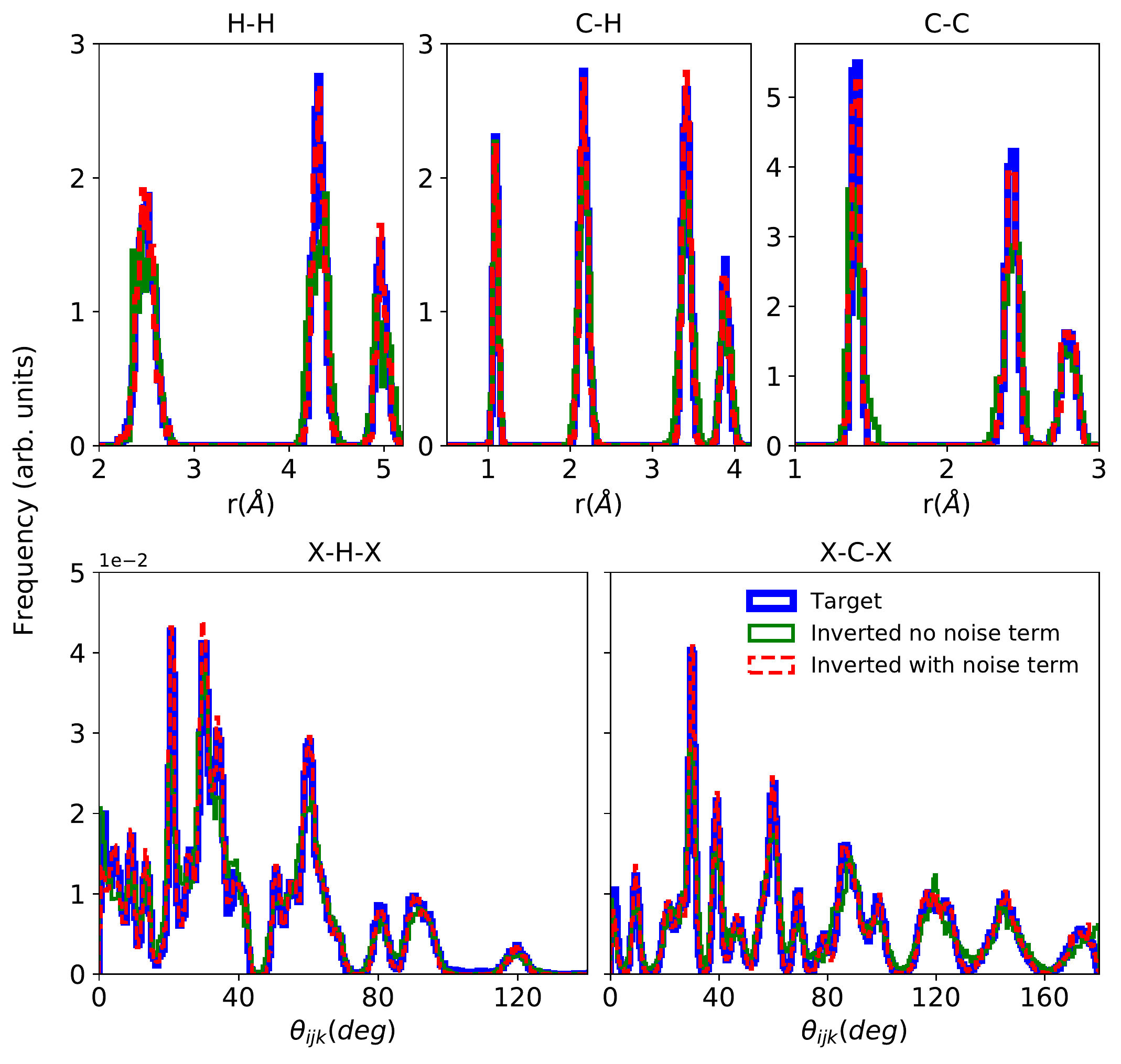}
\caption{(Color on line) Comparison of the partial pair distributions and of the angular distributions between the target 
molecules and the ones resulting from the inversion for a sample of 120 benzene molecules.}
\label{fig:benzene_inversion}
\end{figure}

\begin{figure}
\centering
\includegraphics[scale=0.40]{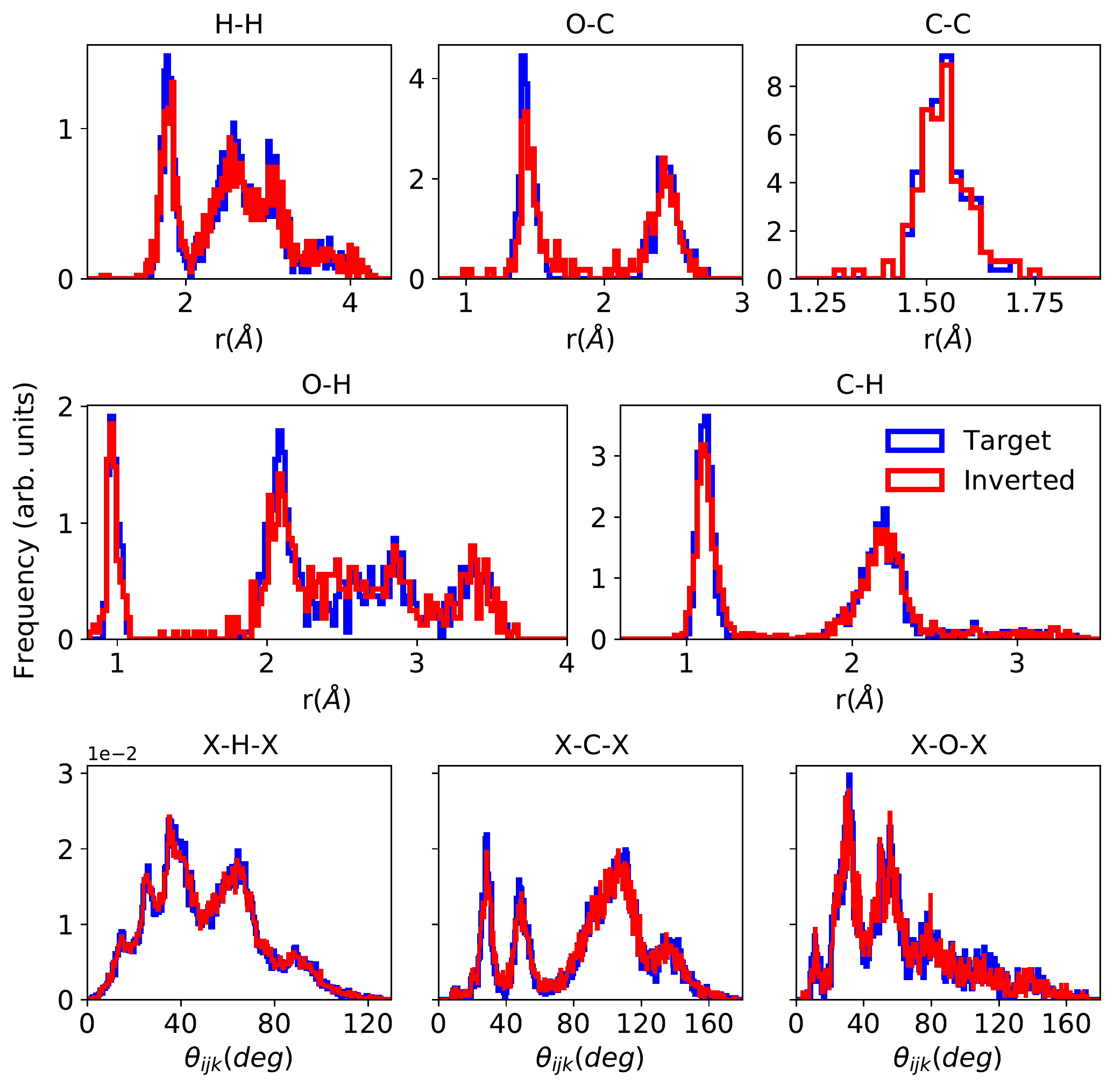}
\caption{(Color on line) Comparison of the partial pair distributions and of the angular distributions between the target 
molecules and the ones resulting from the inversion for a sample of 120 ethanol molecules.}
\label{fig:ethanol_inversion}
\end{figure}
We now proceed to demonstrate the validity of our method by inverting the relation between Cartesian coordinates 
and descriptors for a selected sample of molecules. In particular, here we choose as structural descriptors the bispectrum 
components.\cite{Bartok} In brief, the information about the chemical environment contained within a cutoff radius, 
$r_\mathrm{c}$, from the $i$-th atom can be expressed in term of a neighbour-density distribution,
\begin{equation}\label{density}
\rho_{i}(\mathbf{r})=\delta(\mathbf{r})+\sum_{j} w_{j} \,f_\mathrm{c}(r_{ij})  \, \delta(\mathbf{r}-\mathbf{r}_j)\:,
\end{equation}
where $f_\mathrm{c}$ is a function of the distance $r_{ij}$ between the atoms $i$ and $j$ that smoothly vanishes at 
$r_\mathrm{c}$, while $\{w_{j}\}$'s are weights depending on the species of the $j$-th atom. Then, $\rho_{i}(\mathbf{r})$
can be projected onto the 3-sphere and expanded over hyperspherical harmonics, $U_{m,m'}^{l}$,
%
%
with expansion coefficients, $u^{l}_{m,m'}$. It can be shown\cite{Bartok} that the triple product of expansion 
coefficients, 
\begin{multline}\label{bispectrum}
B_{l_1,l_2,l} = \sum_{m_1,m'_1=-l_1}^{l_1}\sum_{m_2,m'_2=-l_2}^{l_2}\sum_{m,m'=-l}^{l} \left( u^{l}_{m,m'} \right)^*\\ C^{l,l_1,l_2}_{m,m_1,m_2}C^{l,l_1,l_2}_{m',m'_1,m'_2}  u^{l_1}_{m_1,m'_1} u^{l_2}_{m_2,m'_2}\:, 
\end{multline}
is rotationally invariant, where $C^{l,l_1,l_2}_{m,m_1,m_2}$ are the Clebsch-Gordan coefficients.\cite{Varshalovich} 
Such expansion, the bispectrum, contains terms up to the four-body order, if one counts also the central atom.

In figure~\ref{fig:inversion_scheme} we report an example of inversion of the bispectrum components of a deformed 
benzene molecule. The starting configuration in this case is an optimized benzene molecule at equilibrium, with 
a carbon-carbon and a carbon-hydrogen distance of 1.40~\AA\ and 1.09~\AA, respectively. The atomic positions 
are iteratively updated so to minimize the loss of Eq.~(\ref{inversion_loss}).  In order to quantitatively demonstrate 
the quality of the inversion, we compare the atom-pair distances between the molecule before and after the optimization 
procedure against the target. The same comparison is then repeated with the angles formed by all the possible atoms
triplets in the molecule. For both quantities we observe that $5\cdot10^{4}$ iterations are enough to give a molecule 
closely resembling the target one. In Fig.~\ref{fig:inversion_scheme} we also show results for an inversion obtained 
without the noise term in the update rule, Eq.~(\ref{update_rule}). In this second case the final configuration reaches 
a local minimum of $L(\mathbf{r})$ and all atoms remain constrained to the planar arrangement of the initial configuration, 
even though the target has atoms located out of plane. This is because the gradients of the bispectrum components in 
the out-of-plane direction of a planar configuration are zero. Notably, this occurrence is not restricted to the case of 
planar molecules, but it appears at configurational high-symmetry points. The noise term included in the update rule 
breaks all the possible symmetries of the initial configuration, thus improving the outcome of the inversion procedure. 

For a more systematic study we use as target descriptors the bispectrum components of a sample of molecules 
extracted from the MD-17 benchmark dataset,\cite{gdml} containing {\it ab-initio} molecular dynamics trajectories 
of simple organic molecules at 500~K. By taking the first relaxed geometry of the molecular dynamics trajectory
as initial configuration, we now use the inversion algorithm to generate from the bispectrum the Cartesian coordinates 
of $120$ target configurations. The parameters used for the gradient descent algorithm are 
$\gamma=4\times10^{-8}\textup{\AA}^2$,$\,\eta=1\times10^{-2}\textup{\AA}$ and $\nu=1\times10^{-3}$, while the 
bispectrum components have been computed with: $l_\mathrm{max}=4$ and $r_\mathrm{c}=6.0\,\textup{\AA}$.
In Fig.~\ref{fig:benzene_inversion} we report, for benzene, the partial pair-distance distributions and the angular 
distributions of the inverted molecules compared with the targets after $5\times10^{4}$ iterations. Clearly, our inversion 
procedure appears able to generate configurations, which closely reproduce the structural distributions of the targets. 
Notably, the exclusion of the noise term in the update rule leads to a deterioration in performance of the inversion, as 
shown previously.
\begin{figure}
\centering
\includegraphics[scale=0.48]{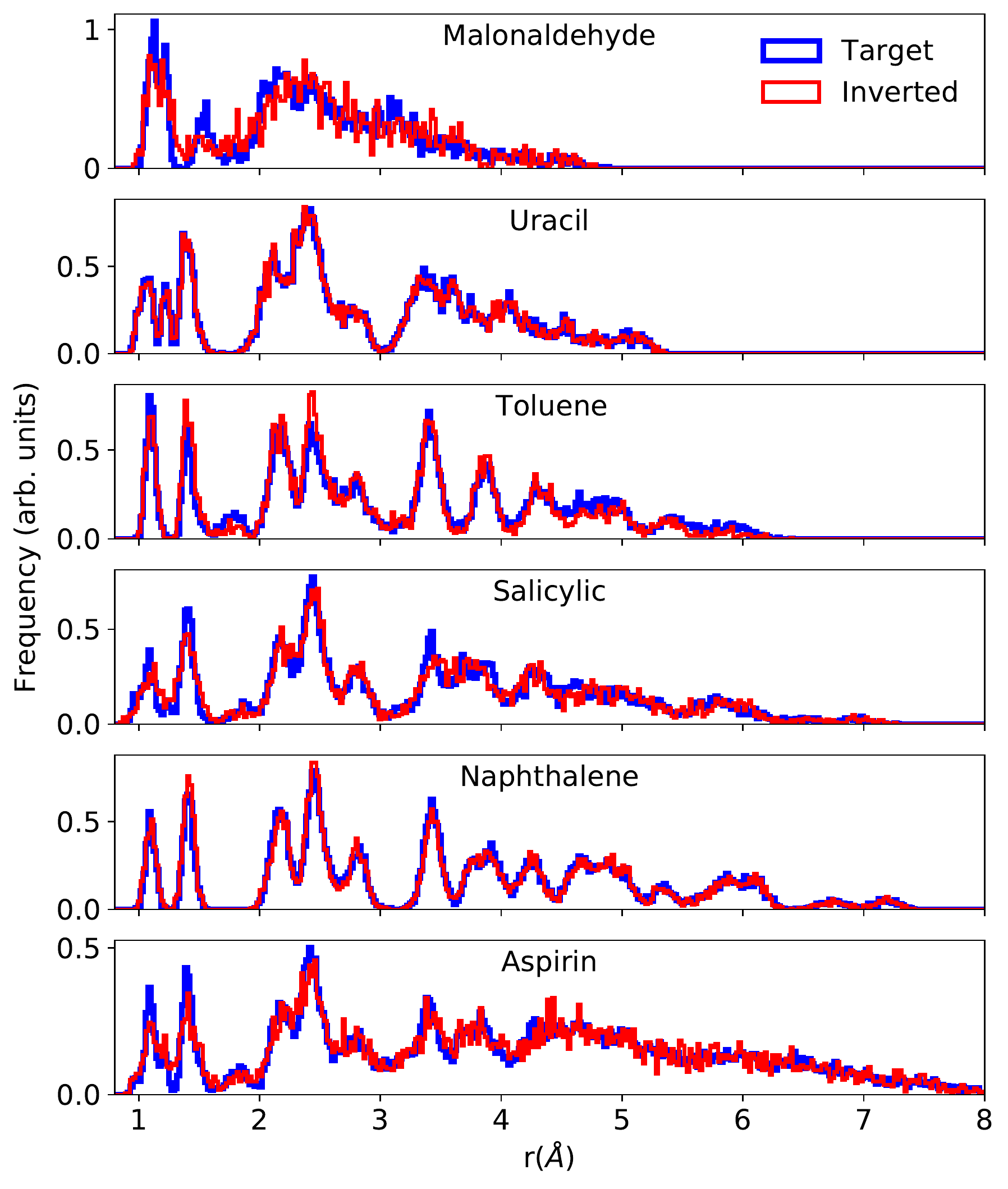}
\caption{(Color on line) Comparison of the total pair distribution between the target molecules and the ones resulting 
from the inversion for a range of molecules contained in the MD17 dataset.}
\label{fig:m17_inversion}
\end{figure}

Benzene represents an optimal choice for the application of our inversion procedure, since atoms of the same specie 
are all equivalent and overall the molecule is fairly rigid. Thus, a harder test is offered by ethanol, where each carbon 
atom is immersed in a different local chemical environment and both the C-O and C-C bonds are mobile, thus allowing 
for molecule torsions. Inversion results are reported in Fig.~\ref{fig:ethanol_inversion} for a sample of $120$ ethanol 
molecules, where again we compare the partial pair and the angular distributions between the inverted configurations 
and the targets. Also in this case, the distributions associated to the inverted configurations are in good agreement 
with the targets. In particular, the width of the peaks in the partial pair distributions are very similar. In this second 
example, however, we find a few configurations to be more problematic to invert given a certain initial configuration. 
This leads to a general deterioration in the inversion procedure, which is reflected in the different peaks heights of 
the two distributions. Such less-converged configurations can be identified by the loss alone, since during the optimisation
$L(\mathbf{r})$ stops improving and stabilises at a value relatively higher than the converged one, thus at a local minimum. 
Multiple restarts of the inversion, with different starting configurations, would lead to certain improvement of the overall 
performance of the algorithm. For instance, the choice of the initial configuration could be driven by the knowledge of 
the vibrational modes of the molecule. We have not explored this route here, and the same starting configuration has 
been used for each inversion.

Finally we test the inversion process over the remaining molecules of the MD-17 dataset, with results reported in 
Fig.~\ref{fig:m17_inversion}. An inspection of the loss evolution with the iteration number for Malonaldehyde, Salicylic 
acid and Aspirin shows that several configurations reach a local minimum of the loss function, resulting only in a partial 
optimisation of the molecule and leading to an incomplete ability to reproduce the target distributions. In contrast, for
Uracil, Toluene and Naphthalene almost all configurations are converged as confirmed by the remarkable similarity 
between the two distributions. 


In summary, we have introduced a universal method that allows one to reconstruct the Cartesian coordinates 
associated to a given set of many-body structural descriptors. This has been implemented with the bispectrum 
components, but can be combined with any local descriptors, such as the power spectrum \cite{Bartok} or 
symmetry functions.\cite{Behler-Parrinello} We have demonstrated the performance of the method over a 
range of simple molecules and shown that the algorithm can efficiently overcome local minima by introducing 
a noise term in the iteration updating rule. Further improvement can be achieved by performing the simulations
over a multiple set of initial conditions. As it stands, our method is the only one available to date to invert many-body
representations back to Cartesian and can be used as a platform in generative methods for molecules/solids design.

{\it Acknowledgement:} MC and PC thank the Irish Research Council for financial support. We acknowledge the 
DJEI/DES/SFI/HEA Irish Centre for High-End Computing (ICHEC) and Trinity Centre for High Performance Computing 
(TCHPC) for the provision of computational resources. The code used in this work is available at 
\url{https://github.com/MCobe94/descriptors-inversion}. It makes use of LAMMPS \cite{lammps} for the computation 
of the bispectrum components, DASK\cite{dask} for scalability and the Atomic Simulation Environment (ASE) 
\cite{ase1,ase2} for the manipulation of the atomic structures.


\begin{thebibliography}{hbp}

\bibitem{Behler-Parrinello}J. Behler and M. Parrinello, 
{\it Generalized Neural-Network Representation of High-Dimensional Potential-Energy Surfaces},
Phys. Rev. Lett. {\bf 98}, 146401 (2007).

\bibitem{Gap}A.P. Bart\'ok, M.C. Payne, T. Kondor and G. Cs\'anyi, 
{\it Gaussian Approximation Potentials: The Accuracy of Quantum Mechanics, without the Electrons},
Phys. Rev. Lett. {\bf 104}, 136403 (2010).

\bibitem{snap}A.P.~Thompson, L.P.~Swiler, C.R.~Trott, S.M.~Foiles and G.J. Tucker,
{\it Spectral neighbor analysis method for automated generation of quantum-accurate interatomic potentials},
J. Comp. Phys. {\bf 285}, 316 (2015).

\bibitem{ANI}J. Smith, O. Isayev and A. Roitberg, 
{\it ANI-1: An extensible neural network potential with DFT accuracy at force field computational cost},
Chem. Sci. {\bf 8}, 3192 (2017).

\bibitem{gdml}S. Chmiela, A. Tkatchenko, H.E. Sauceda, I. Poltavsky, K.T. Sch{\"u}tt, K.-R. M{\"u}ller, 
{\it Machine learning of accurate energy-conserving molecular force fields},
Science Adv. {\bf 3}, e1603015 (2017).

\bibitem{Lunghi2019a}A. Lunghi and S. Sanvito, 
{\it A unified picture of the covalent bond within quantum-accurate force fields: from simple organic molecules to metallic complexes' reactivity}, 
Science Advances {\bf 5}, eaaw2210 (2019).

\bibitem{Lunghi2019b}A. Lunghi and S. Sanvito, 
{\it Surfing multiple conformation-property landscapes via machine learning: Designing magnetic anisotropy}, 
J. Chem. Phys. C {\bf 124}, 5802 (2019).

\bibitem{Musil2021}F. Musil, A. Grisafi, A.P. Bart\'ok, C. Ortner, G. Cs\'anyi and M. Ceriotti,
{\it Physics-Inspired Structural Representations for Molecules and Materials},
Chem. Rev. {\bf 121}, 9759 (2021).

\bibitem{Bartok}A.P. Bart\'ok, R. Kondor and G. Cs\'anyi, 
{\it On representing chemical environments},
Phys. Rev. B {\bf 87}, 184115 (2013).

\bibitem{Sanchez-Lengeling2018}B. Sanchez-Lengeling and A. Aspuru-Guzik,
{\it Inverse molecular design using machine learning: Generative models for matter engineering},
Science {\bf 361}, 360 (2018).

\bibitem{smiles1}D. Weininger, 
{\it SMILES, a chemical language and information system. 1. Introduction to methodology and encoding rules},
J. Chem. Inf. Comput. Sci. {\bf 28}, 31 (1988).

\bibitem{smiles2}W. Wenhao and C.W. Coley, 
{\it The Synthesizability of Molecules Proposed by Generative Models},
J. Chem. Inf. Mod. {\it 60}, 5714 (2020).

\bibitem{drugan}A. Kadurin, S. Nikolenko, K. Khrabrov, A. Aliper and A. Zhavoronkov, 
{\it druGAN: An Advanced Generative Adversarial Autoencoder Model for de Novo Generation of New Molecules with Desired Molecular Properties in Silico},
Mol. Pharmaceutics {\bf 14}, 3098 (2017).

\bibitem{molgan}N. De Cao and T. Kipf,
{\it MolGAN: An implicit generative model for small molecular graphs},
arXiv:1805.11973 (2018).

\bibitem{crystal_gan_1}S. Kim, J. Noh, G.H. Gu, A. Aspuru-Guzik, Y. Jung, 
{\it Generative Adversarial Networks for Crystal Structure Prediction},
ACS Cent. Sci. {\bf 6}, 1412 (2020).

\bibitem{crystal_gan_2}A. Nouira, N. Sokolovska and J.-C. Crivello,
{\it CrystalGAN: Learning to Discover Crystallographic Structures with Generative Adversarial Networks},
AAAI Spring Symposium: Combining Machine Learning with Knowledge Engineering (2019).

\bibitem{crystal_gan_3}Y. Zhao, M. Al-Fahdi, M. Hu, E.M.D. Siriwardane, Y. Song, A. Nasiri and J. Hu,
{\it High-throughput discovery of novel cubic crystal materials using deep generative neural networks},
Adv. Sci. {\bf 8}, 2100566 (2021).

\bibitem{Bishop}C.M. Bishop, {\it Pattern Recognition and Machine Learning (Information Science and Statistics)},
Springer-Verlag, Berlin, Heidelberg (2006).

\bibitem{Varshalovich}D. A. Varshalovich, A. N. Moskalev, V. K. Khersonskii {\it Quantum Theory of Angular Momentum}, World Scientific (1988)

\bibitem{Incompleteness}S.N.~Pozdnyakov, M.J.~Willatt, P.A.~Bart\'ok, C.~Ortner, G.~Cs\'anyi and M.~Ceriotti, 
{\it Incompleteness of Atomic Structure Representations}, 
Phys. Rev. Lett. {\bf 125}, 166001 (2020).

\bibitem{lammps} A. P. Thompson, H. M. Aktulga, R. Berger, D. S. Bolintineanu, W. M. Brown, P. S. Crozier, P. J. in 't Veld, A. Kohlmeyer, S. G. Moore, T. D. Nguyen, R. Shan, M. J. Stevens, J. Tranchida, C. Trott, S. J. Plimpton, 
{\it LAMMPS -  a flexible simulation tool for particle-based materials modeling at the atomic, meso, and continuum scales},  
Comp. Phys. Comm. {\bf 271}, 10817 (2022).

\bibitem{dask} Dask Development Team (2016). Dask: Library for dynamic task scheduling
https://dask.org

\bibitem{ase1} A.H.~Larsen, J.J.~Mortensen, J.~Blomqvist, I.E.~Castelli, R.~Christensen, M. Du\l{}ak, J.~Friis, M.N.~Groves, 
B.~Hammer, C.~Hargus, E.D.~Hermes, P.C.~Jennings, P.B.~Jensen, J.~Kermode, J.R.~Kitchin, E.L.~Kolsbjerg, J.~Kubal, K.~Kaasbjerg,
S.~Lysgaard, J.~Bergmann Maronsson, T.~Maxson, T.~Olsen, L.~Pastewka, A.~Peterson, C.~Rostgaard, J.~Schi{\o}tz, O.~Sch\"utt, 
M.~Strange, K.S.~Thygesen, T.~Vegge, L.~Vilhelmsen, M.~Walter, Z.~Zeng, K.W.~Jacobsen,
{\it The Atomic Simulation Environment: a Python library for working with atoms}, 
J. Phys.: Condens. Matter. {\bf 29}, 273002 (2017).

\bibitem{ase2} S.R. Bahn and K.W. Jacobsen,
{\it An object-oriented scripting interface to a legacy electronic structure code}, 
Comput. Sci. Eng. {\bf 4}, 56 (2002).

\end{thebibliography}
\end{document}